\documentclass[aps,prl,twocolumn,superscriptaddress,showpacs,longbibliography]{revtex4-2}
\usepackage{amsfonts,amsmath,bm}
\usepackage{mathtools}
\usepackage{graphicx}
\usepackage{epstopdf}

\usepackage{amsmath}

\DeclarePairedDelimiter\norm{\lVert}{\rVert}%
\DeclarePairedDelimiter\abs{\lvert}{\rvert}%
\makeatletter
\let\oldabs\abs
\def\abs{\@ifstar{\oldabs}{\oldabs*}}
\let\oldnorm\norm
\def\norm{\@ifstar{\oldnorm}{\oldnorm*}}
\makeatother

\begin{document}

	\title{Near-field plasmonics for generation of phonon lasing in a thermal nanomachine }
	
	\author{P. Karwat}
	\affiliation{School of Physics and CRANN Institute, Trinity College Dublin, Dublin 2, Ireland \\}
	\affiliation{Department of Theoretical Physics, Wroc\l{}aw University of Science and Technology, Wybrze\.ze Wyspia\'nskiego 27, 50-370 Wroc\l{}aw, Poland\\}
	\author{G. Pasławski}
	\affiliation{Department of Theoretical Physics, Wroc\l{}aw University of Science and Technology, Wybrze\.ze Wyspia\'nskiego 27, 50-370 Wroc\l{}aw, Poland\\} 
	\author{P. Damery}
	\affiliation{School of Physics and CRANN Institute, Trinity College Dublin, Dublin 2, Ireland \\}
	\author{Y. Yang}
	\affiliation{School of Physics and CRANN Institute, Trinity College Dublin, Dublin 2, Ireland \\}
	\author{F. Bello}
	\affiliation{School of Physics and CRANN Institute, Trinity College Dublin, Dublin 2, Ireland \\}  
	\author{O. Hess}
	\affiliation{School of Physics and CRANN Institute, Trinity College Dublin, Dublin 2, Ireland \\} 
	\affiliation{The Blackett Laboratory, Department of Physics, Imperial College London, South Kensington Campus, SW7 2AZ, London, United Kingdom \\} 
	\vspace{1cm}
	\date{\today}
	
	\begin{abstract}
		Recent advances in near-field plasmonic metamaterials, such as  nanoresonators or transducers, have demonstrated the ability to generate localized fields of high intensity, and thus maintain relatively large nanoscale heat gradients on the order of $10^1-10^2$~K/nm. A plasmonic near-field transducer (NFT) can achieve such large gradients, making population inversion achievable within phononic media. We herein develop a thermal nanomachine composed of a nanoscale phononic laser using InGaAs quantum dot media where an NFT serves as the plasmonic energy source. We show, on demand, the generation of phonons while having full control of the phonon lasing medium. We also demonstrate the ability to obtain population inversion of a photonic transition in the system if one chooses. 
	\end{abstract}
	
	\pacs{Valid PACS appear here}
	\maketitle
	\section{\label{sec:introduction}I. Introduction}
	Lasers not only play an important role in a number of scientific fields such as optics and metrology \cite{LasersApplications,Cruz:10}, but they are a crucial part of many devices used in our daily lives. The optical properties they offer are well known~\cite{Gu:98,Ojambati:21,McKenna:21}, but there is much more to discover. Herein, we utilize a very similar model to that of a coherent photon laser in order to create a phenomenon of phonon occurrence, i.e. phonon lasing. Phonons offer a wide range of wavelengths not accessible to conventional lasers while opening up novel methods for communication devices. We achieve this by using a 3-level system coupled to heat baths, `hot and cold', thus creating a heat gradient throughout the structure. Due to the heat gradient, we are able to observe excitations that lead to an inversion of the occupation of phonon states.
	\begin{figure}[h!]
		\begin{center}
			\includegraphics[width=80mm]{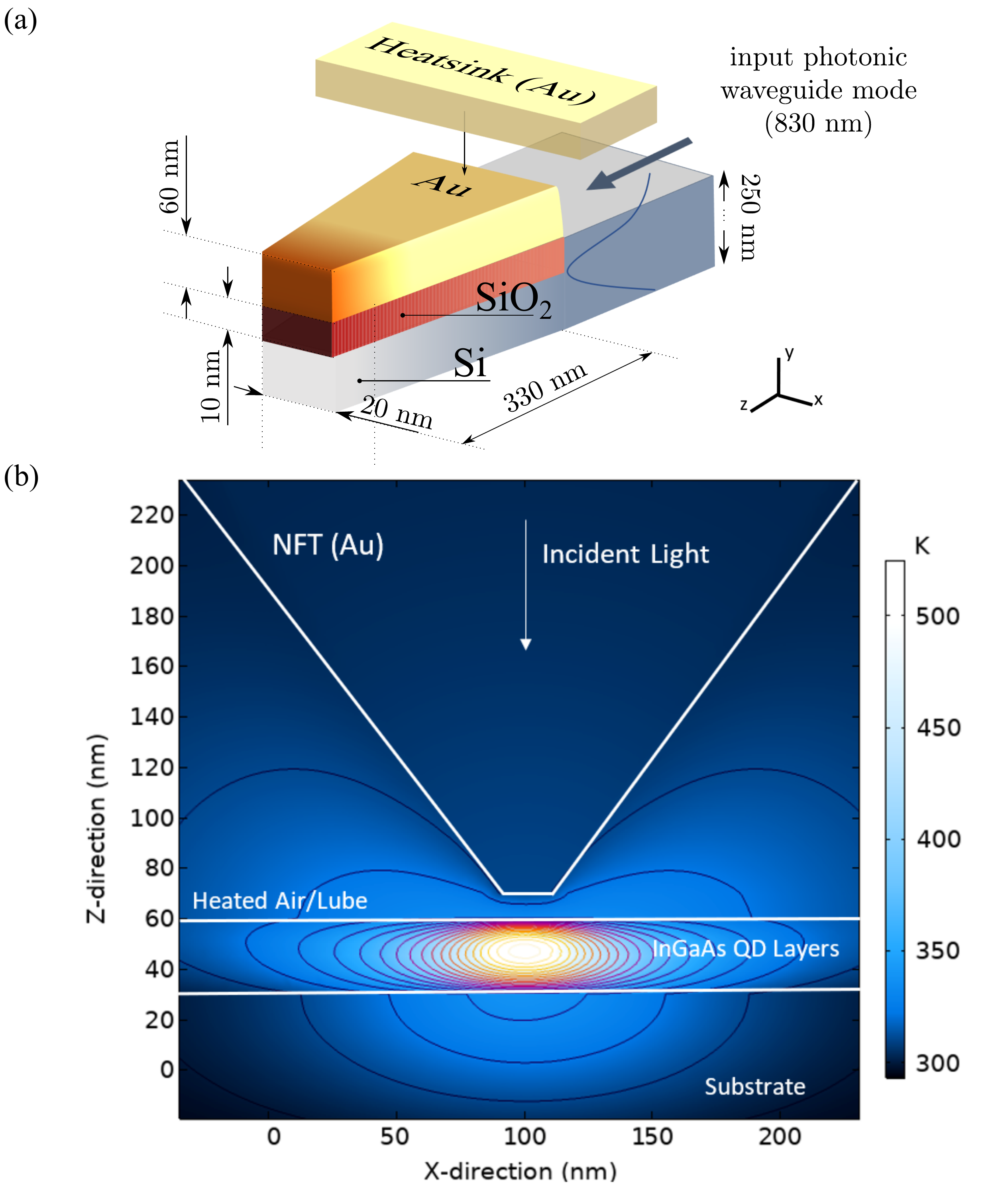}
		\end{center}
		\caption{\label{fig:NFT}(Color online) (a) Minimal model of the plasmonic near-field transducer (NFT) where light of 830 nm couples to a plasmon mode which propagates along the $\mathrm{Au-SiO}_{2}$ interface \cite{Abadia:18,Bello:19}. An Au heatsink is added to keep the operating temperature of the Au well under 400 K to prevent protrusion due to thermal expansion \cite{yates2013thermal,Richter:13}. (b) Temperatures can exceed well over 500 K for an input power of 0.9375 mW (shown) and over 700 K if it is doubled. Contour lines follow the region of strongest temperature gradients necessary for the nanoscale phonon laser we describe with gradients approximate to state-of-the-art values within present-day experiments (see supplementary information for further details).}  
	\end{figure}
	
\begin{figure}[h!]
		\begin{center}
			\includegraphics[width=85mm]{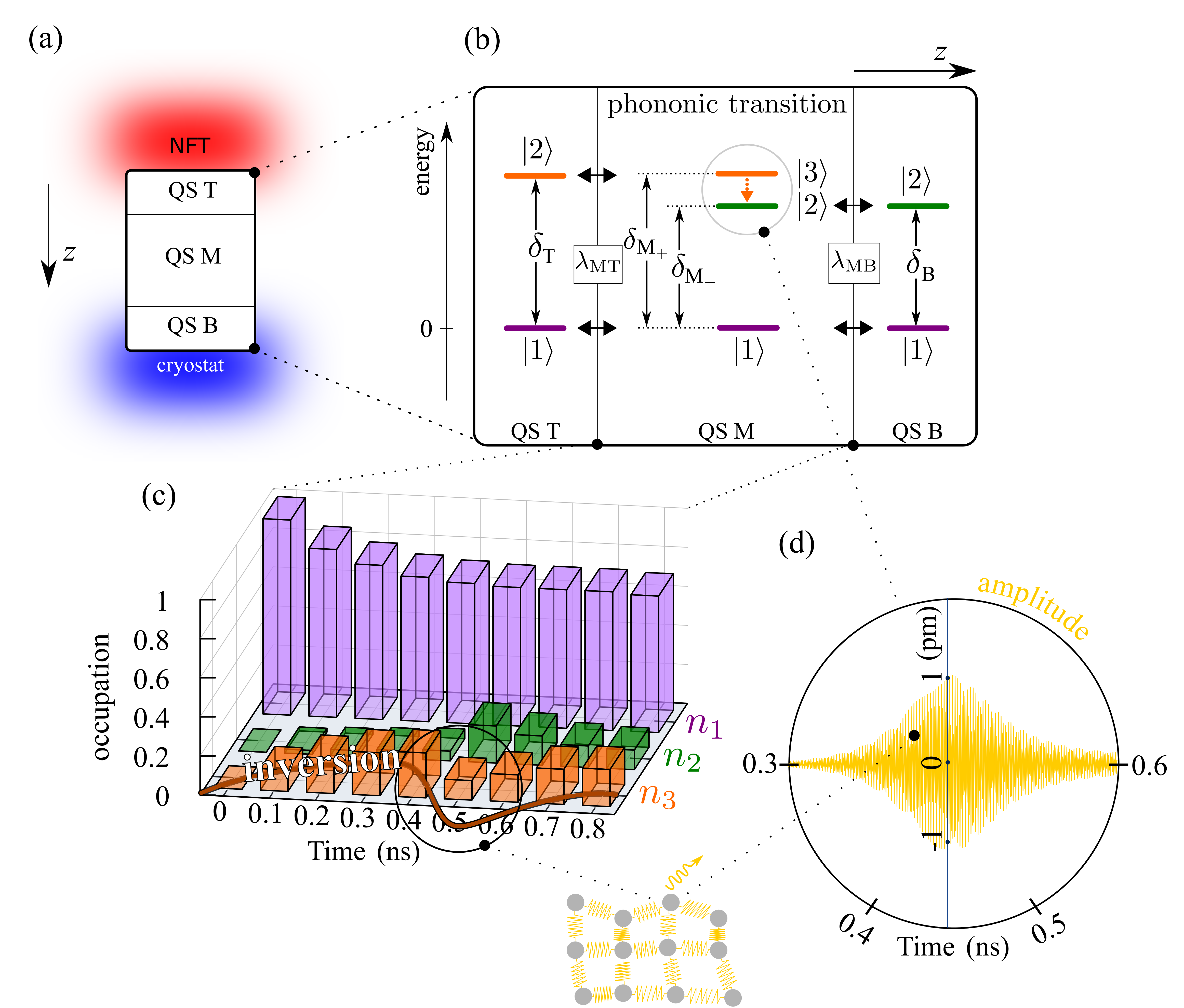}
		\end{center}
		\caption{\label{fig:3lvl_intro}(Color online) Phonon lasing medium. (a) A sketch of the system with the NFT at the top and cryostat at the bottom. (b) Energy levels of quantum subsystems with energy mismatch parameter $\Delta = \delta_{\mathrm{T}} - \delta_{\mathrm{M_{+}}}$ and a pair of two upper states highlighted in the middle subsystem (QS M) coupled to phonon reservoir. QS T refers to the subsystem on top and QS B is the bottom subsystem. (c) Population probability in QS M with the crucial inversion ($n_{3} - n_{2}$) plotted. The colors of the 3D bars refer to the states shown above. (d) Full dependence of lattice displacement field on time with schematically added emission of a phonon.}  
	\end{figure}

	We focus on a nanoscopic thermal system consisting of three coupled subsystems as depicted in Fig.~\ref{fig:3lvl_intro}; a hot bath, a 3-level system for inversion, and a cold bath. Our study determines how the system would behave depending on the temperature difference and strength of coupling between the hot bath and 3-level subsystems. The detailed analysis uncovers new possibilities for the generation of lasing using acoustic wavelengths of phonons, along with a better understanding of phonon lasing systems on the nanoscale and how to build a device using near-field plasmonics~\cite{Li2021,Zograf:21}.

	\section{\label{sec:system}II. System}
	
	The thermal nanomachine used to generate a coherent source of phonons is built of three quantum subsystems and two heat baths coupled together and created by a plasmonic near-field transducer (NFT). The~top (quantum-subsystem top, QS T) and bottom (QS B) subsystems – with two energy levels – work as filters, converting incoherent input (energy) into coherent output, and also preventing the accumulation of thermal occupation, and allowing the inversion to happen. The middle one (QS M) – with three energy levels – is the active medium used to create inversion for phonon lasing. The third energy level in the QS M initially has the same value as the upper state of QS T, while the second energy level in QS M is in resonance with the upper state of the QS B. The lower energy level is set to be equal in all subsystems.
	
	The heat gradient we use may be constructed, and adeptly manipulated, on the nanoscale using recent advances in near-field plasmonics. In particular, to create the relatively large temperature change necessary very strong heat gradients on the order of many Kelvins per nanometer~\cite{Richter:13} are required. There are a few plasmonic metamaterials, e.g. resonators, proposed that could manifest such gradients with only recent reports showing them achievable even in academic labs~\cite{Halas:21}. Our work investigates the thermal gradients induced from a non-integrated, plasmonic NFT, which subdiffracts a photonic mode by coupling it to a surface plasmon mode of the metal-insulator-semiconductor NFT shown in Fig.~\ref{fig:NFT} \cite{Abadia:18,Bello:19}. Crucially, this design is non-integrated with the 3-level subsystem, and serves to manipulate temperatures of the heat baths. Dependent on the absorbing media, the NFT creates the large heat gradients required for phonon lasing. The gradient ultimately couples the QSs to heat baths with high ($T_\mathrm{H}$) and low ($T_\mathrm{C}$) temperatures. In the examined case, we neglect interactions between any non-adjacent parts of the device.
	Furthermore, the interactions between subsystems can be controlled by changing the energies of the states.
	
	Due to the flow of excitations caused by heat gradient, we observe the emission of coherent phonons from QS M as a result of the achieved inversion between states 3 and 2 in the subsystem. In order to describe this system properly, we have to write its Hamiltonian as a sum of three elements. First, one must describe the energies of each state in all three subsystems, reading
	\begin{equation}
		\hat{H}\mathrm{_{sys}} = \hat{h}\mathrm{^{(T)}} + \hat{h}\mathrm{^{(M)}} + \hat{h}\mathrm{^{(B)}}.
	\end{equation}
	These elements are all expanded in analogous fashion, using $\epsilon$ to describe the energies and $\hat{P}$ as standard projection operators, i.e.
	\begin{equation}
		\hat{h}\mathrm{^{(T)}} = \sum_{\mathrm{n}}^2 \epsilon\mathrm{_n^{(T)}}\hat{P}\mathrm{_{nn}^{(T)}} \otimes \hat{1}\mathrm{^{(M)}} \otimes \hat{1}\mathrm{^{(B)}}.
	\end{equation}
	The component of the total Hamiltonian that describes interactions between adjacent subsystems, which depends on the coupling strength ($\lambda\mathrm{_{MT/MB}}$), reading
	\begin{equation}
		\hat{H}\mathrm{_{int}} = \lambda\mathrm{_{MT}} \bigg(\hat{h}\mathrm{^{(TM)}} \otimes \hat{1}\mathrm{^{(B)}}\bigg) + \lambda\mathrm{_{MB}}\bigg(\hat{1}\mathrm{^{(T)}} \otimes \hat{h}\mathrm{^{(MB)}}\bigg),
	\end{equation}
	with the coupling
	\begin{equation}
		\hat{h}\mathrm{^{(TM)}} = \hat{P}^{\mathrm{(T)}}_{21} \otimes \bigg(\hat{P}^{\mathrm{(M)}}_{12} + \hat{P}^{\mathrm{(M)}}_{13} + \hat{P}^{\mathrm{(M)}}_{23}\bigg) + \mathrm{h.c.}
	\end{equation}
	and analogous for $\hat{h}\mathrm{^{(MB)}}$.
	
	Taking into account the interactions between the middle subsystem and the phonon reservoir with coupling coefficient $g$ and dephasing rate $\Gamma$, we use values of $u_0$ as the initial amplitude for the lattice displacement and $u^{(+/-)}$ derived from the ensemble average, 
	\begin{equation*}
		\langle \hat{u} \rangle = u^{(-)} + u^{(+)},	
	\end{equation*}
	we get
	\begin{equation}
		\hat{H}\mathrm{_{ph}} = \hbar \omega \hat{b}^{\dagger} \hat{b} + \hat{1}\mathrm{^{(T)}} \otimes \frac{\hbar g}{u_0}(u^{(-)} \hat{P}\mathrm{_{23}^{(M)}} + u^{(+)} \hat{P}\mathrm{_{32}^{(M)}}) \otimes \hat{1}\mathrm{^{(B)}},
	\end{equation}
	where $\hat{b}^{\dagger} (\hat{b})$ are the raising (lowering) operators for the phonon field. After that, we use a semi-classical master equation in Lindblad form to predict time evolution of the whole system, given by
	\begin{equation}
		\frac{d\hat{\rho}}{dt} = -\frac{i}{\hbar}\big[\hat{H}\mathrm{_{sys}} + \hat{H}\mathrm{_{int}} + \hat{H}\mathrm{_{ph}},\hat{\rho}\big] + \hat{D}\mathrm{_{H}}(\hat{\rho}) + \hat{D}\mathrm{_{C}}(\hat{\rho}).
	\end{equation}
	
	Elements denoted as $\hat{D}\mathrm{_{H/C}}$ describe the coupling with hot (H) and cold (C) heat baths. Lindblad operators $\hat{L}$ and the effectiveness of the heat coupling given by the rates  $\Gamma\mathrm{_k}$ \cite{breuer2002theory,Scully:97} are dependent on the coupling strength and the energy difference between states yielding,
	\begin{equation}
		\hat{D}\mathrm{_H}(\hat{\rho}) = \sum_{\mathrm{k}=1}^2 \Gamma\mathrm{_k}(T\mathrm{_H}) \big( \hat{L}\mathrm{_k^{(T)}}\hat{\rho}\hat{L}\mathrm{_k^{(T)\dagger}} - \frac{1}{2}\{\hat{L}\mathrm{_k^{(T)\dagger}}\hat{L}\mathrm{_k^{(T)}},\hat{\rho}\} \big)
	\end{equation}
	and analogous results for $\hat{D}\mathrm{_C}$.
	
	The equation of motion for the lattice displacement then reads
	\begin{equation}
		\frac{du}{dt} = - \Gamma u + iu_0g\big(\rho\mathrm{_{23}^{(M)}}(t) - \rho\mathrm{_{32}^{(M)}}(t)\big)
		\label{LD}
	\end{equation}
	
	\noindent with $\rho\mathrm{_{mn}}$ being the expectation value of the projection operator evaluated using the trace of the density matrix.
	
	\begin{figure}[t]
		\begin{center}
			\includegraphics[width=70mm]{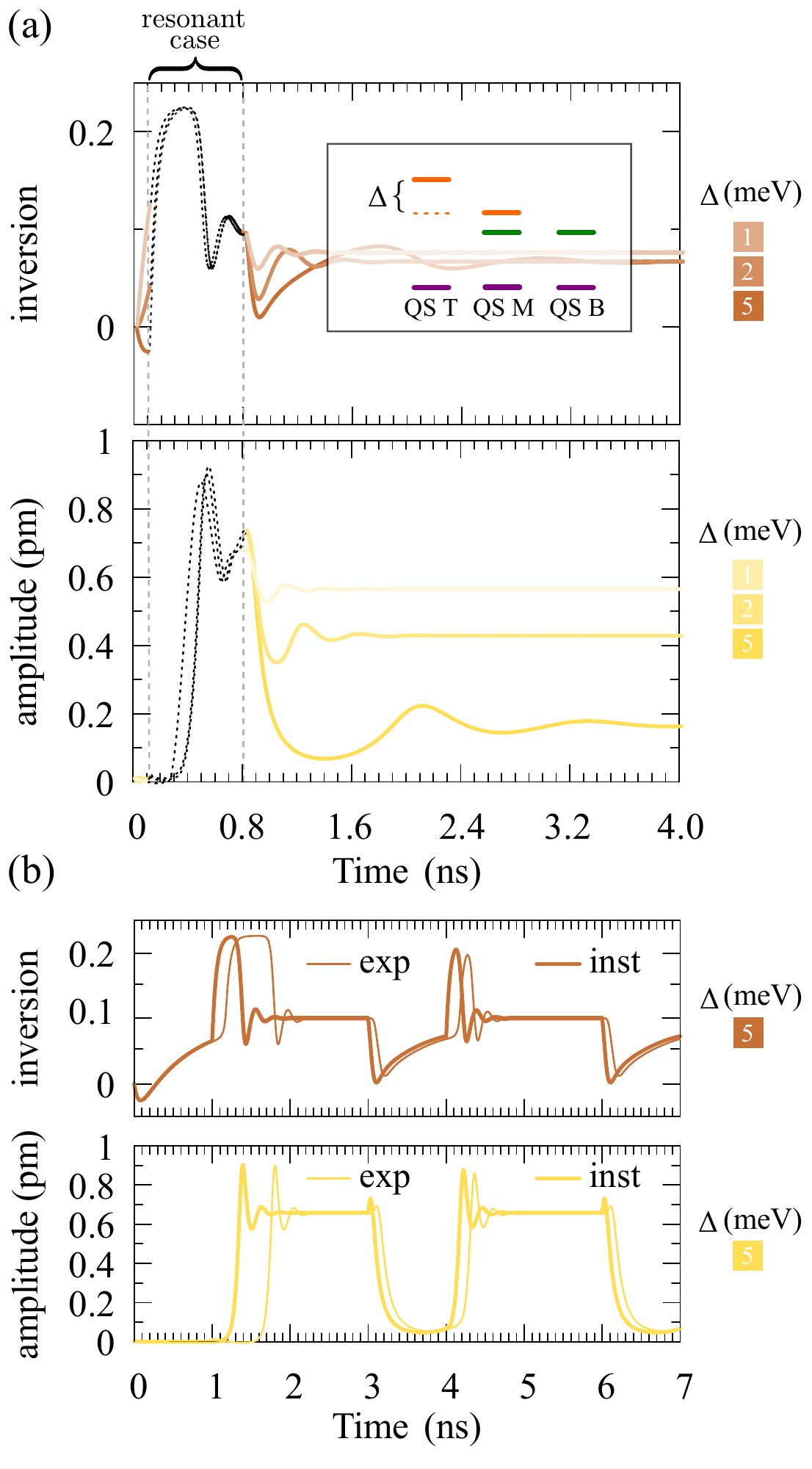}
		\end{center}
		\caption{\label{fig:inv_off_res}(Color online) (a) Taking control of the system by breaking the resonant condition; from the beginning to 0.1~ns and over 0.8~ns (the darker the color, the stronger energy mismatch). (b) Comparison of switching the system on and off instantly (thick solid lines) and exponentially (fine solid lines); for inversion of occupations (top) and amplitude of lattice displacement field (bottom) in both cases.} 
	\end{figure}
	For our simulations (if not stated otherwise) we assumed $\delta\mathrm{_{M+}} = 30$~meV, $\delta\mathrm{_{M-}} = 25$~meV, $\gamma = 3$~ps$^{-1}$, $\Gamma = 2$~ps$^{-1}$, $g = 2.25$~ps$^{-1}$, $u_{0} = 20$~pm. Temperature given by the transducer is set to $T_\mathrm{NFT} = 400$~K, while the cryostat $T_\mathrm{C} = 100$~K.
	\section{\label{sec:results}III. Controlling of phonon lasing}
	
	Demonstrating coherent control is fundamental for practical applications of the phonon laser, especially in the nanoworld. One possibility for achieving this is to break the resonance between subsystems. Parameter $\Delta$ in Fig.~\ref{fig:inv_off_res} (inserted graph) depicts the energy mismatch between the second level of QS T and the third level of QS M.\\ 
	In Fig.~\ref{fig:inv_off_res}(a), we take a closer look at how the off-resonant condition affects the results. The energy of QS~T is placed off resonance initially (from 0 to $t = 0.1$~ns) with $\Delta$ far away from the denoted value. Then, it is instantly switched on to the nearly resonant case ($\delta_{\mathrm{T}} \approx \delta_{\mathrm{M+}})$. Then at $t = 0.8$~ns, it was set back to the initial condition again. As expected, changing the energy mismatch leads to the deterioration of coupling strength. Here, we observe an impact on oscillations of the lattice displacement field resulting in final vaule of the steady-state, which is due to reduced inversion in QS M. Only for $\Delta=0$ we get the most efficient flow of excitation (not shown).\\
	In reality, switching is always accompanied by a time delay. In Fig.~\ref{fig:inv_off_res}(b), we compare instant switching (thick solid line) with a more slower, exponential change (fine solid line), for a chosen energy mismatch $\Delta = 5$~meV. The shape of both curves allows the corresponding peaks to match, but it can be seen that an exponential change gives a smoother curve without rapid soaring or diving of values for inversion and amplitude, respectively.
	\begin{figure}[t]
		\begin{center}
			\includegraphics[width=80mm]{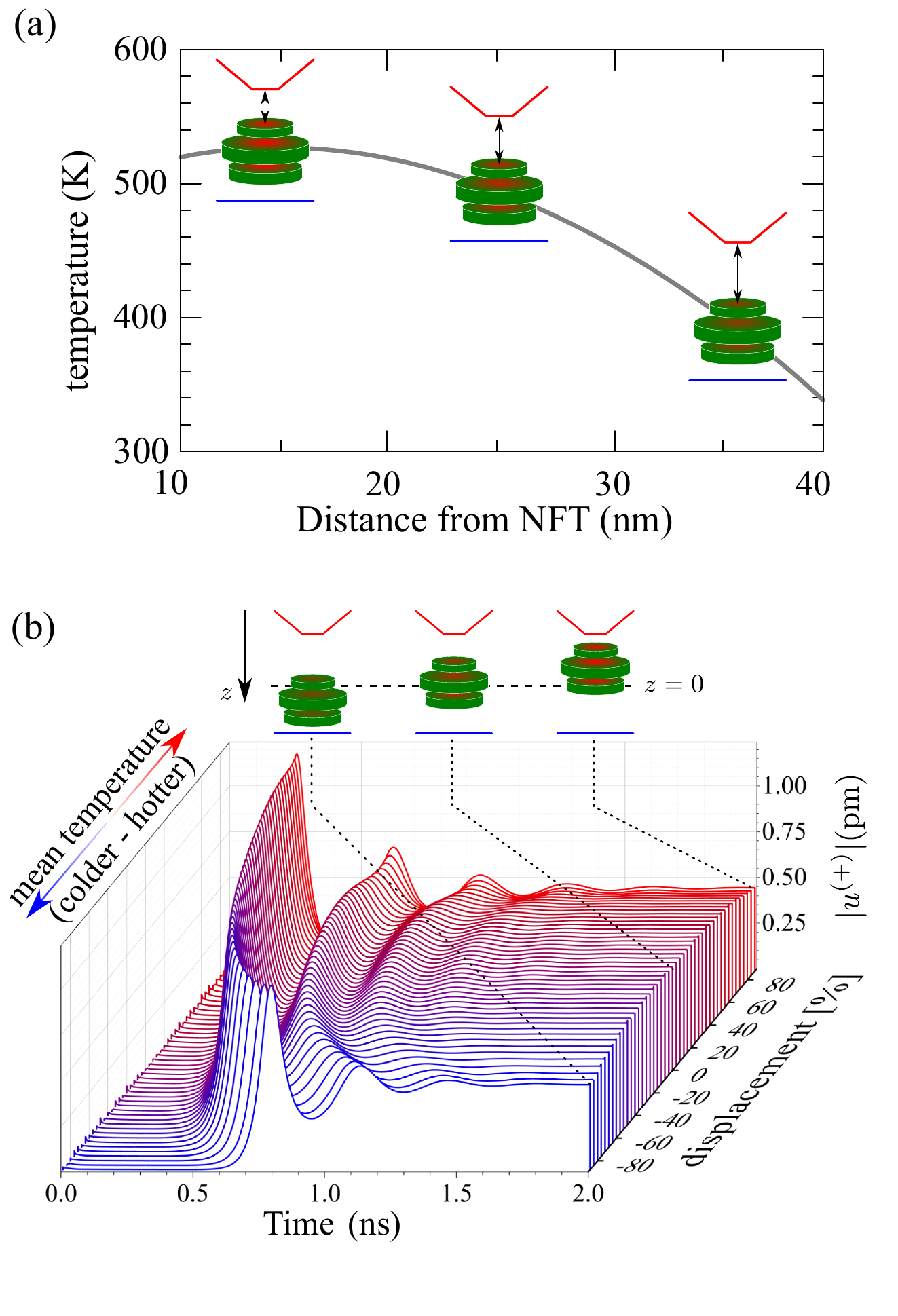}
		\end{center}
		\caption{\label{fig:heat}(Color online) (a) Temperature at the surface of QS~T as a function of the change in distance of the QDs to the NFT. (b) Amplitude of lattice displacement field as a function of time and the amount of displacement (percentage) of the QDs to the hotter or colder heat bath with the mean temperature depicted.}  
	\end{figure}
	
	Recently, we have shown the impact of the mean temperature of the system on the inversion~\cite{Karwat:18}. In this work, we take a further step to consider spatial dependence. The NFT is a source of energy needed for the flow of excitation.\\ 
	In Fig.~\ref{fig:heat}(a) we show the temperature dependence of the NFT position relative to the QD layer, while the distance to the colder cryostat is fixed. One can observe the significant decrease of surface temperature with the change in distance (from several to dozens of nm). In Fig.~\ref{fig:heat}(b) we plot a set of curves that represent the temperature change for a given amount of displacement. At the beginning the system is placed in the middle, i.e. $z=0$ (having the same distance to top and bottom heat baths). Then, we moved the system closer to the hotter or colder one, in the range from 0 to $\pm80\%$ of the whole distance between them. The mean temperature of the system is depicted as well. As a result, the colder the mean temperature, the longer the time needed for the amplification of the lattice displacement field. We note that such a system might be represented by a set of vertically stacked self-assembled QDs \cite{Karwat:15} embedded within a nanowire \cite{Kats:11,Heiss:13}, with different size and content, while we operate on the familiar energy scale.   
	
	\begin{figure}[t]
		\begin{center}
			\includegraphics[width=90mm]{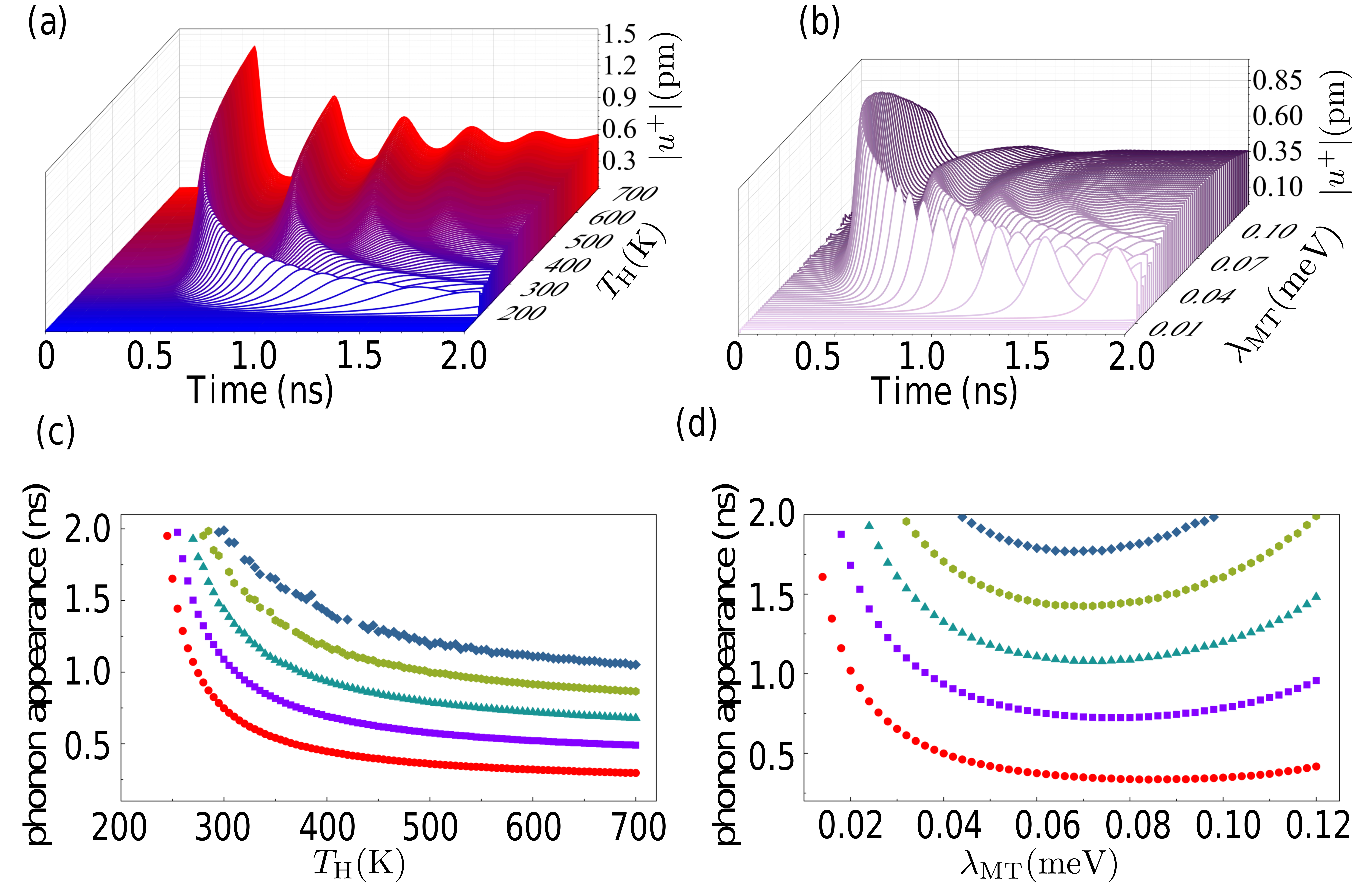}
		\end{center}
		\caption{\label{fig:peaks}(Color online) Lattice displacement field for (a) temperature dependence for time evolution of the system. (b) coupling strength dependence for time evolution of the system. (c) peak location with respect to the temperature for five first maxima of the lattice displacement field's amplitude where exponential behaviour can be identified; (d) peak location with respect to the coupling strength between the top and middle subsystems for the first five maxima of the lattice displacement field's amplitude. Their positions reveal stretched, yet parabolic behaviour.}  
	\end{figure}
	
	Next, we investigate how the system behaves depending on a wide range of parameters. Specifically, we have chosen to examine the temperature difference and coupling between top and middle subsystem ($\lambda\mathrm{_{MT}}$). First, we focus on the temperature dependence. While all parameters were kept constant (including $T_\mathrm{C}$), the simulations were run for $T_\mathrm{H}$ in the range from 100~K to 700~K with a 5~K step. Fig.~\ref{fig:peaks}(a) shows the lattice displacement field for each iteration. It can easily be seen that not only the amplitude grows with higher temperature, but also the frequency of relaxation oscillations increases significantly. The shape arranges itself into sets of successive peaks and valleys indicating that there are oscillations in the system after the first phonon is generated. Their amplitude decreases with time and a~steady state is reached. We also analyze the location of peaks in this chart by connecting the maximum of the amplitude with the time needed for its appearance. This leads to a~series of curves resembling nearly exponential functions (Fig.~\ref{fig:peaks}(c)). That would suggest there is a~threshold indicating the minimal time before the first phonon appearance. On the other hand, the time necessary for phonon occurrence will grow rapidly with temperature difference decreasing.
	A second analysis has been done considering the change of coupling strength between the top and middle subsystems. This time, the temperature of the hotter bath $T_\mathrm{H}$ is set to 400~K for all simulations, while $\lambda\mathrm{_{MT}}$ was examined in the range from 0.002~meV to 0.12~meV with a 0.002~meV step.
	The results show that we are not able to set as high a value of $\lambda\mathrm{_{MT}}$ as we want and expect the outcome to get better as we move towards stronger coupling. Even though the amplitude is not strongly affected by the change, we consider the impact of coupling in greater detail.
	Nearly parabolic behaviour of the valleys is easily spotted in Fig.~\ref{fig:peaks}(d); as the minimal time needed for the phonon appearance decreases with coupling strength until it reaches value around 0.07~meV. Then it starts increasing again, as the coupling strength grows. It is worth noting that the location of the vertex of these assumed parabolas is not in the same place for all acquired maximums. They are slighty shifted towards lower values of $\lambda\mathrm{_{MT}}$ and may be seen for higher order oscillations. It is also worth noting that each consecutive valley tends to be less "flat", and therefore we surmise that for some sets of parameters we could acquire a wide range of values for the coupling strength that assures fast phonon generation for at least first two phonon occurances.\\
	
	\section{\label{sec:resultsB}IV. Conclusion and Comments on Two-Pulse Propagation}
\begin{figure}[t]
		\begin{center}
			\includegraphics[width=90mm]{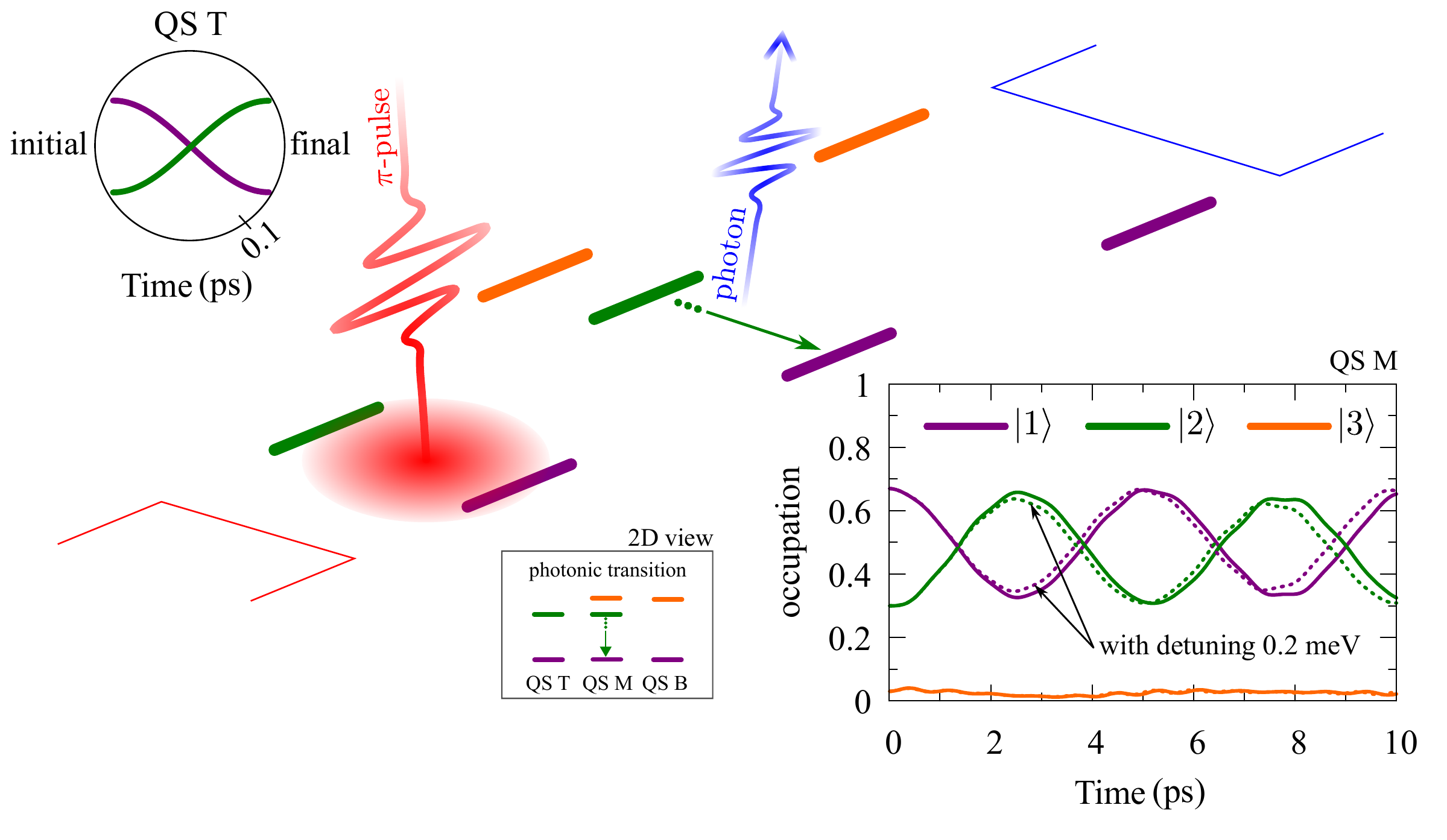}
		\end{center}
		\caption{\label{fig:pipulse}(Color online) Sketch of another configuration of the system to allow the emission of photon. The $\pi$-pulse is used to swap occupations in QS T at very short time scale (tenths of ps), which is faster than the time of oscillating inversion we observe in QS M (a few ps).} 
	\end{figure}

	We propose a very interesting scenario that assumes two-pulse propagation where we can potentially control the emission of either phonon-photon, phonon-phonon, or photon-photon pairs depending on the lasing medium. Here, we aim to show that population inversion of photonic transitions is also possible in our level system which we now excite using a laser pulse.
	Modifying Eq.~\eqref{LD} to consider laser excitation, we get the equation for the electric field amplitude
	\begin{equation}
		\frac{dE}{dt} = - \gamma_{\mathrm{res}} E + iCM\big(\rho\mathrm{_{12}^{(M)}}(t) - \rho\mathrm{_{21}^{(M)}}(t)\big),
		\label{EFA}
	\end{equation}
	where $\gamma_{\mathrm{res}}$ is the loss rate of the resonator and $C$ is the coupling between light and the gain material response, $M$ is dipole matrix element.
	This allows us to consider a photonic transition from state $|2\rangle$ to $|1\rangle$ as shown in Fig.~\ref{fig:pipulse} that is coupled to the electric field. We use a $\pi$ - pulse to act on the two-level filter (calculations done in accordance with Quantum Jump Method \cite{molmer_93}), i.e. the left subsystem (QS T), to swap the occupation probabilities of each level. Indeed, the coupling between transitions 1 and 2 of the left (QS T) and middle (QS M) subsystems, which are slightly detuned, then allows the photonic transition to QS M (inserted graph). At the moment, decay rates are neglected to demonstrate proof-of-concept though they are anticipated to be on the picosecond order for plasmonically-coupled QDs that include enhanced spontaneous emission \cite{Akselrod:14}. We note that the population inversion of the photonic transition may also be achieved if the two-level system of QS T is on resonant with the NFT (830 nm), thus making the external  $\pi$ - pulse obsolete. Therefore, the NFT could potentially achieve inversions and/or strong coupling with both the phonon and photon transitions in the system \cite{Bello:20}. This is penned for future investigation.
	
	To conclude, we have presented the properties of a~heat-gradient driven nanomachine with a plasmonic NFT as a non-integrated source of thermal energy, which allows the system to realize and maintain phononic population inversion. We have shown the generation of coherent phonons on demand as a~pulsed source. The description was based on a~Lindblad form of the quantum master equation with a~semiclassical coupling to a~single phonon mode, i.e. lattice displacement field. We have also made a~first step towards the spatial analysis and the mean temperature of the system by taking a~closer look at how close the NFT must be located to the QD layer in order to realize the crucial inversion and, in consequence, the generation of coherent phonons. Remarkably, the system shows an amplification of the lattice amplitude for a~wide range of parameters checked. Having this, we can start to think about possible implementation of phonon lasing medium on a~semiconductor system or vibrational molecules. This includes the combination of phononic and photonic lasing. Moreover, further investigation of these properties could give insight into the nature of phonon lasing using heat gradients, and as future work quantum heat transport shall be taken into consideration \cite{Pekola:21,Cui:19}.
	
	\vspace{2cm}
	\acknowledgments
	We would like to thank prof. Jeremy Baumberg for discussion and advice, and prof. Doris E. Reiter for inspiring questions. This work was funded by the Science Foundation of Ireland
	(SFI) Grant 18/RP/6236.
	\newpage
	\bibliographystyle{prsty}
	\bibliography{reference}

\begin{thebibliography}{10}

\bibitem{LasersApplications}
J.~J. Wilson, {\em Lasers : principles and applications}, {\em Prentice Hall
  International series in optoelectronics} (Prentice Hall, New York ; London,
  1987).

\bibitem{Cruz:10}
F.~C. Cruz,  in {\em Latin America Optics and Photonics Conference} (Optical
  Society of America, United States, 2010), p.\ WH1.

\bibitem{Gu:98}
B. Gu and I. Franco, Phys. Rev. A {\bf 98},  063412  (2018).

\bibitem{Ojambati:21}
O.~S. Ojambati, K.~B. Arnardottir, B.~W. Lovett, J. Keeling, and J.~J.
  Baumberg, Few-emitter lasing in single ultra-small nanocavities, 2021.

\bibitem{McKenna:21}
R. McKenna, S. Corbett, S.~T. Naimi, D. Mickus, D. McCloskey, and J.~F.
  Donegan, IEEE Journal of Selected Topics in Quantum Electronics {\bf 28},  1
  (2022).

\bibitem{Abadia:18}
N. Abad\'{i}a, F. Bello, C. Zhong, P. Flanigan, D.~M. McCloskey, C. Wolf, A.
  Krichevsky, D. Wolf, F. Zong, A. Samani, D.~V. Plant, and J.~F. Donegan, Opt.
  Express {\bf 26},  1752  (2018).

\bibitem{Bello:19}
F. Bello, S. Sanvito, O. Hess, and J.~F. Donegan, ACS Photonics {\bf 6},  1524
  (2019).

\bibitem{yates2013thermal}
B. Yates, {\em Thermal Expansion}, {\em Monographs in Low-Temperature Physics}
  (Springer US, United States, 2013).

\bibitem{Richter:13}
H.~J. Richter, C.~C. Poon, G. Parker, M. Staffaroni, O. Mosendz, R. Zakai, and
  B.~C. Stipe, IEEE Transactions on Magnetics {\bf 49},  5378  (2013).

\bibitem{Li2021}
Y. Li, W. Li, T. Han, X. Zheng, J. Li, B. Li, S. Fan, and C.-W. Qiu, Nature
  Reviews Materials {\bf 6},  488  (2021).

\bibitem{Zograf:21}
G.~P. Zograf, M.~I. Petrov, S.~V. Makarov, and Y.~S. Kivshar, Adv. Opt. Photon.
  {\bf 13},  643  (2021).

\bibitem{Halas:21}
P.~D. Dongare, Y. Zhao, D. Renard, J. Yang, O. Neumann, J. Metz, L. Yuan, A.
  Alabastri, P. Nordlander, and N.~J. Halas, ACS Nano {\bf 15},  8761  (2021),
  pMID: 33900744.

\bibitem{breuer2002theory}
H. Breuer and F. Petruccione, {\em The Theory of Open Quantum Systems} (Oxford
  University Press, Oxford, 2002).

\bibitem{Scully:97}
M. Scully and M. Zubairy, {\em Quantum Optics} (Cambridge University Press,
  United Kingdom, 1997).

\bibitem{Karwat:18}
P. Karwat, D.~E. Reiter, T. Kuhn, and O. Hess, Phys. Rev. A {\bf 98},  053855
  (2018).

\bibitem{Karwat:15}
P. Karwat and P. Machnikowski, Phys. Rev. B {\bf 91},  125428  (2015).

\bibitem{Kats:11}
V.~N. Kats, V.~P. Kochereshko, A.~V. Platonov, T.~V. Chizhova, G.~E. Cirlin,
  A.~D. Bouravleuv, Y.~B. Samsonenko, I.~P. Soshnikov, E.~V. Ubyivovk, J.
  Bleuse, and H. Mariette, Semiconductor Science and Technology {\bf 27},
  015009  (2011).

\bibitem{Heiss:13}
M. Heiss, Y. Fontana, and A. Gustafsson, Nature Mater {\bf 12},  439  (2013).

\bibitem{molmer_93}
K. Mølmer, Y. Castin, and J. Dalibard, Journal of the Optical Society of
  America B {\bf 10},  524  (1993).

\bibitem{Akselrod:14}
G.~M. Akselrod, C. Argyropoulos, T.~B. Hoang, C. Cirac{\`i}, C. Fang, J. Huang,
  D.~R. Smith, and M.~H. Mikkelsen, Nature Photonics {\bf 8},  835  (2014).

\bibitem{Bello:20}
F. Bello, N. Kongsuwan, J.~F. Donegan, and O. Hess, Nano Letters {\bf 20},
  5830  (2020).

\bibitem{Pekola:21}
J.~P. Pekola and B. Karimi, Rev. Mod. Phys. {\bf 93},  041001  (2021).

\bibitem{Cui:19}
L. Cui, S. Hur, and Z. Akbar, Nature Mater {\bf 12},  439  (2013).

\end{thebibliography}

\end{document}